\documentstyle[epsfig]{aipproc}

\begin{document}
\hfill RAL-TR-97-003
\vspace{-0.1in}
\title{A Leading--Order, But More Than One--Loop, 
\\ Calculation of Structure Functions}

\author{Robert S. Thorne}
\address{Rutherford Appleton Laboratory\\
Chilton, Didcot\\Oxon, U.K.}

\maketitle
\vspace{0.2in}
\centerline{\bf To be published in the proceedings of DIS97, 
Chicago, April 1997.}

\vspace{-0.2in}

\begin{abstract}
I present a full leading--order calculation of 
$F_2(x,Q^2)$ and $F_L(x,Q^2)$, including contributions not only from 
leading order in $\alpha_s$, but also from the leading power of 
$\alpha_s$ for each order in $\ln(1/x)$. The calculation is ordered  
according to the inputs
and evolution of the structure functions, and the perturbative 
form of the inputs is determined. I compare the results of fits to data to 
those using conventional LO and NLO order calculations, and the correct 
inclusion of leading $\ln(1/x)$ terms is clearly preferred. A prediction for 
$F_L(x,Q^2)$ is produced which is smaller at small $x$ than 
that obtained from the conventional approach.  
\end{abstract}

\section*{Introduction}

There has recently been a great deal of new data obtained at HERA 
for the structure function $F_2(x,Q^2)$ at small $x$ \cite{hone,zeus}, 
and consequently a great 
deal of theoretical activity. The main theoretical question 
is whether one should include leading $\ln(1/x)$ terms which 
cause small $x$ enhancement of terms at high orders in $\alpha_s$, and if 
so, then in precisely what manner. Common wisdom was that such terms should 
not be included because standard ways of doing so showed 
that the terms were indeed important for the ranges of $x$ and $Q^2$ 
probed at HERA, but that they worsened global fits to data obtained
from conventional LO and NLO approaches, rather than 
improved them. Indeed, the best global fits to data seemed to come from 
NLO evolution
starting at $Q_0^2\sim 2\hbox{\rm GeV}^2$ where the input for the singlet 
quark distribution (weighted by $x$), and hence the structure function,  
behaved like $x^{-0.25}$ at small $x$ \cite{rome}. 
Thus, despite the good quality 
of the global fits, the situation was unsatisfactory since
terms were ignored because they were inconvenient, rather than 
for any sound theoretical reason, and the required input for the 
quark distribution was of a rather steep, unjustified form.  

Recently I proposed that a correct theoretical treatment of the 
calculation of structure functions demands that the leading 
powers of $\alpha_s$ for given powers of $\ln(1/x)$ must be included, but  
in expressions for physical quantities\cite{thorne}. This latter point was 
inspired by, and is similar to the 
idea of using physical anomalous dimensions proposed by Catani
\cite{catani}. However, the structure imposed by 
an ordered expansion within a given renormalization scheme 
has a number of other consequences for the form of the expressions for 
structure functions. I will discuss the  
leading--order--renormalization--scheme--consistent (LORSC) calculation of
structure functions briefly before looking at comparison with experiment.

\section*{The LORSC Calculation.}

The principles underlying the calculation and the consequences are:

\medskip

\noindent 1. The quantities one calculates to a given order are
directly observable. Hence rather than coefficient functions and parton 
distributions the structure function is factorized into inputs at
some scale $Q_I^2$ and evolution away from this scale. One obvious 
consequence of this is factorization scheme independence. 

\medskip

\noindent 2. For a physical quantity the leading--order 
expression for each independent component begins at its lowest power of 
$\alpha_s$, whatever power that happens to be, e.g. if 
a term of $\ln^m(1/x)$ appears for the first time at order 
$\alpha_s^n$ this is the 
leading--order term for $x$-dependence of this type. In practice this is 
implemented in moment space where the leading power in $\alpha_s$ for each
inverse power of the moment variable $N$ is part of the full 
leading--order expression. 
As it must, this results in leading--order expressions 
which are renormalization scheme independent and which are therefore 
compatible with the use of the one--loop running coupling. 

\medskip

\noindent 3. The inputs for the structure functions are two fundamental 
nonperturbative functions, one each for $F_2$ and $F_L$, convoluted with
perturbative contributions. The nonperturbative functions are taken to be flat 
at small $x$. The perturbative parts are determined so that the total 
expression for the structure function is independent of the choice of 
starting scale, i.e. the LORSC expression is invariant under changes 
of $Q_I^2$ up to changes beyond leading order. This determines the 
inputs uniquely up to a scale, $Q^2_{NP}$, which should be  
typical of the onset of nonperturbative physics. 
For both $F_2$ and $F_L$ the inputs
are required to behave roughly like $x^{-0.28}$ for $Q_I^2\sim 20-100 
\hbox{\rm GeV}^2$ for $10^{-2} \leq x \leq 10^{-5}$.
Moreover, the forms of $F_L(Q^2_I)$, $F_2(Q^2_I)$ and 
$(d\,F_2(Q^2)/d\,\ln Q^2)_{Q^2_I}$ are all
related at small $x$, and once one of them is set (in practice 
$F_2(Q^2_I)$) by fitting to data there is very little freedom in the others, 
a constraint largely absent in conventional approaches.  

\medskip

These points are discussed in far greater detail in \cite{thorne}, and
the way in which they are put into practice is presented in the latter of 
these papers. 

\section*{Fits to Data and Predictions}

The fit using the LORSC expressions is performed with the parameters 
specifying the form of the 
nonperturbative inputs and the nonperturbative scale left free, 
and is repeated for a wide variety of input scales $Q_I^2$. The best fit 
comes from $Q^2_I=40 \hbox{\rm GeV}^2$, but is very insensitive to $Q^2_I$
for values from $20 \hbox{\rm GeV}^2 \to 80 \hbox{\rm GeV}^2$. We find
that $Q^2_{NP}= 0.55 \hbox{\rm GeV}^2$, precisely the sort of value expected.
The charm and bottom quarks are dealt with rather naively, treated as massless
above threshold ($4 \hbox{\rm GeV}^2$ and $20 \hbox{\rm GeV}^2$ respectively)
and playing no role below this. The charm threshold is chosen to give a 
reasonable description of data on the charm structure function.
$Q^2$ is chosen as the (squared) renormalization scale, and the value of 
$\Lambda_{QCD}$ is held fixed at $100 \hbox{\rm MeV}$
(giving $\alpha_s(M_Z^2)=0.115$ at one loop), though this is 
certainly close to the best fit value.
The results of the fit are shown in table \ref{table1}.    

The conventional NLO fit is performed in the usual manner and the input scale
for the parton distributions $Q_0^2$ is chosen to be the same as that for the 
charm threshold ($2.75 \hbox{\rm GeV}^2$). For the best fit we find that
$\Lambda_{{\overline {\rm MS}}} =300 \hbox{\rm MeV}$, i.e. 
$\alpha_s(M_Z^2)=0.118$. The input quark distribution 
$\sim x^{-1-0.22}$. The conventional LO fit is performed in the same way with
charm threshold of $3 \hbox{\rm GeV}^2$, $\Lambda_{QCD}=160 \hbox{\rm MeV}$,
(one--loop $\alpha_s(M_Z^2)=0.124$),
and small $x$ quark $\sim x^{-1-0.23}$. The results of these more 
conventional fits are also shown in table \ref{table1}.\footnote{ All fits 
were constrained not only by consistency with charm 
structure function data, but also by consistency with high--$x$ prompt photon 
data.}  

\begin{table}[b!]
\caption[]{Quality of fits using full leading--order (LORSC) 
expressions and conventional LO and NLO expressions.}
\label{table1}
\begin{tabular}{ldddd}
$x$-range& data points& LORSC $\chi^2$& NLO $\chi^2$ & LO $\chi^2$\\
\tableline
$x \geq 0.1$& 551 & 622 & 615 & 605\\
$x<0.1$& 548 & 483 & 554 & 598 \\
total & 1099 & 1105 & 1169 & 1203\\
\end{tabular}
\end{table}

In each case the values of $F_2(x,Q^2)$ used for the HERA data
are not simply those published but are corrected for the values of 
$F_L(x,Q^2)$ predicted in each of the fits. For the LORSC fit this causes a 
slight lowering of $F_2(x,Q^2)$ for some points due to smaller predicted 
$F_L(x,Q^2)$ which in fact improves the fit to the small $x$ data by 
a $\chi^2$ of about 5. 
For the NLO fit there is essentially no correction, but for the LO fit the 
predicted $F_L(x,Q^2)$ is extremely large since the fit requires both 
a large coupling and large gluon at small $x$. 
Hence, the values of $F_2(x,Q^2)$ are 
corrected upwards, sometimes by $4\%$, leading to 
the correct LO fit being $\sim 25$ worse than a fit to 
uncorrected data.\footnote{For the LO and 
NLO fits the normalization of the H1 data is as low as I allow, i.e. 
$98.5\%$, while for the LORSC fit it is $100\%$. In each case the 
normalization of the ZEUS data is $\sim 1\%$ higher.} 

As one can see the LORSC fit is clearly superior: although the 
difference in quality is $\sim 0.06$ per point this is significant
when considering 1099 data points.
Indeed, the difference between the quality of the LORSC fit and the NLO fit
is larger than that between the NLO and LO fits. 
The quality in terms of different experiments may be found in the 
second of\cite{thorne} (as may the references for the experiments), but it is 
illustrative to present it in terms of two bins in $x$.  
There is little difference in the quality of the high--$x$ fits, as
we would expect since at high $x$ the missing $\ln(1/x)$ terms 
in the LO and NLO fits cause no problems (and an appropriate choice of 
coupling makes the LO fit as good as the NLO
fit). In fact, the LORSC calculation
is missing NLO  $\alpha_s$ terms, and should perhaps be a little worse 
than the NLO fit.\footnote{This is clear if we perform fits to 
high--$x$ data only: the LORSC fit only improves by a few points, but 
the NLO fit improves to 578, with
$\Lambda_{{\overline {\rm MS}}} = 215 \hbox{\rm MeV}$. 
Such a low value of $\Lambda_{{\overline {\rm MS}}}$ is not allowed in 
the full NLO fit since it leads to a very poor fit at small $x$.}
However, at small $x$ there is a very clear 
deterioration in the quality of fit going from LORSC to NLO to LO
(which would be even greater if values of 
$\Lambda$ consistent with the best high--$x$ fits were used). This 
is due to the loss of vital $\ln(1/x)$ terms as ones goes from LORSC to NLO, 
and also, to a lesser extent, when going from 
NLO to LO. Hence, the results of the fits are precisely as we would expect 
from theoretical arguments. They clearly imply that in determining values of 
$\alpha_s(M_Z^2)$ and the gluon distribution in particular schemes the 
NLO fit should only be trusted at relatively high $x$.  

\begin{figure}
\centerline{\epsfig{file=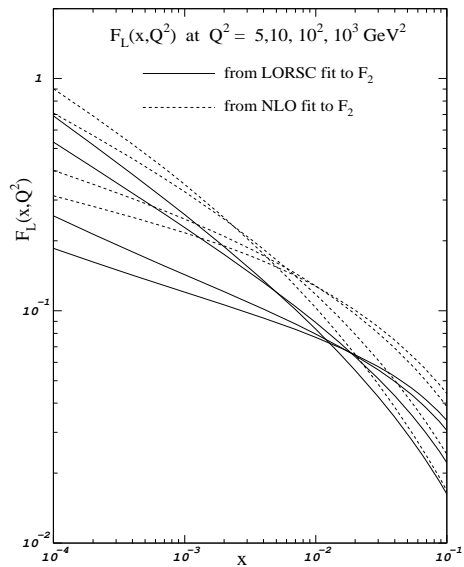,height=3.5in,width=2.5in}}
\vspace{-0.32in}
\caption[]{Comparison of predictions of $F_L(x,Q^2)$ using the LORSC fit and 
the NLO fit. In both cases $F_L(x,Q^2)$ increases with $Q^2$ at small $x$.}
\label{fig}
\end{figure}

Shown in fig. \ref{fig} is the comparison of the predictions for $F_L(x,Q^2)$ 
made using the best fit from the LORSC approach and the NLO approach. 
As one can
see there is significant difference between them, and any relatively 
accurate measurement of $F_L(x,Q^2)$ should see some sign of which
approach is preferred. The ``determination'' of 
$F_L(x,Q^2)$ by H1 is no use in differentiating between the two since it 
is really just a consistency check. If one assumes that the 
H1 NLO fit is correct and extrapolates from low $y=Q^2/xs$, where the 
determination of $F_2(x,Q^2)$ from the cross--section is 
insensitive to $F_L(x,Q^2)$ then the difference between the measured 
cross--section and the extrapolation provides a value of 
$F_L(x,Q^2)$\cite{flong}.
This must be consistent with the NLO prediction for $F_L(x,Q^2)$, which it is. 
However, assuming that the LORSC calculation is correct the
extrapolation is different, and hence so is the determined 
$F_L(x,Q^2)$. Consistency between the determined $F_L(x,Q^2)$ and the 
prediction is also good in this case.
\footnote{The H1 fit results in a gluon which is inconsistent with both 
high--$x$ prompt photon data and moderate--$x$ charm data, and it would be 
illustrative to repeat the process with a more constrained NLO fit. It 
would also be interesting to use all available high $y$ data, i.e. also 
include ZEUS data. Such a study is being performed.} 
Hence, a true measurement of $F_L(x,Q^2)$ would be an invaluable aid to the 
determination of the real physics describing hadron interactions at small 
$x$. An important role may also be played by improved data on the charm 
structure function, and incorporating heavy quarks correctly into the LORSC 
approach is a project well underway. Less inclusive quantities may well
also play an important part, but calculations remain to be done. 

\section*{Conclusion}

A comparison to all significant data gives clear evidence that
the LORSC calculation of structure functions
is preferred to the usual NLO--in--$\alpha_s$ approach. 
Thus, although the usual expansion technique seems to be acceptable at 
small $x$ at present, and will probably be used for most QCD calculations in
the near future, when attempting to describe any data at small $x$ 
it should be borne in mind that this approach may well be untrustworthy: 
anomalies may occur, and
more careful calculations may be needed.  
The situation will be clarified by a variety of different small $x$ 
measurements.


\begin{references}
\bibitem{hone}H1 collaboration, {\it Nucl. Phys.} {\bf B470} 3 (1996). 
\bibitem{zeus}ZEUS collaboration, {\it Zeit. Phys.} {\bf C69} 607 (1996);
ZEUS collaboration, {\it Zeit. Phys.} {\bf C72} 399 (1996).
\bibitem{rome}Ball, R.D., and De Roeck, A., Summary Talk of WGI at DIS96,
Rome, April 199, and references therein.
\bibitem{thorne}Thorne, R.S., {\it Phys. Lett.} {\bf B392} 463 (1997);
{\tt hep-ph/9701241}, preprint RAL-TR-96-065, Jan 1997. 
\bibitem{catani}Catani, S., talk at UK workshop on HERA physics, September
1995, unpublished; {\tt hep-ph/9609263}, preprint DDF248/4/96, April 1996;
Proc. of DIS96, Rome, April 1996.
\bibitem{flong}H1 collaboration, {\it Phys. Lett.} {\bf B393} 452 (1997).
\end{references}
\end{document}